\documentclass[lettersize,journal]{IEEEtran}
\usepackage{amsmath,amsfonts}
\usepackage{algorithmic}
\usepackage{algorithm}
\usepackage{array}
\usepackage{textcomp}
\usepackage{stfloats}
\usepackage{url}
\usepackage{verbatim}
\usepackage{graphicx}
\usepackage{cite}
\usepackage[table]{xcolor}
\usepackage{amsthm}
\usepackage{amssymb}
\usepackage{graphicx}
\usepackage{float}
\usepackage{balance}
\usepackage{enumerate}
\usepackage{color}
\usepackage{array}
\usepackage{indentfirst}
\usepackage{multirow}
\usepackage{booktabs}
\usepackage{dsfont}
\usepackage{comment}  
\usepackage{mathrsfs}
\usepackage{xcolor}
\usepackage{hyperref}
\usepackage{bbm}
\usepackage[normalem]{ulem} 

\usepackage{xcolor}
\usepackage{caption}
 
\usepackage[caption=false,font=footnotesize]{subfig} 
\usepackage[font=footnotesize,labelsep=period]{caption}
 
\newtheoremstyle{IEEEitalic}%
  {\topsep}   
  {\topsep}   
  {\normalfont}  
  {2em}            
  {\itshape}    
  {:}           
  {.5em}        
  {}            
 
\theoremstyle{IEEEitalic}

\renewcommand \thetable{\Roman{table}}
\renewcommand{\figurename}{Fig.}
 
\DeclareCaptionLabelFormat{blue-label}{\textcolor{blue}{#1~#2}}

\makeatletter

\newcommand{\Rmnum}[1]{\expandafter\@slowromancap\romannumeral #1@}
\makeatother

\definecolor{revised}{RGB}{0 0 0}
\definecolor{question}{RGB}{0 0 0}

\usepackage{titlesec}
\titlespacing*{\section}
  {2pt}{1ex plus .2ex minus .1ex}{0.6ex plus .1ex}
\titlespacing*{\subsection}
  {0pt}{0.8ex plus .2ex minus .1ex}{0.4ex plus .1ex}

\begin{document}

\title{Diffusion Model-Enhanced Environment Reconstruction in ISAC}

\author{Nguyen Duc Minh Quang,
        Chang Liu,~\IEEEmembership{Member,~IEEE},
        Shuangyang Li,~\IEEEmembership{Member,~IEEE},
        Hoai-Nam Vu,
        
        Derrick Wing Kwan Ng,~\IEEEmembership{Fellow,~IEEE},
        Wei Xiang,~\IEEEmembership{Senior Member,~IEEE}
\thanks{N.~D.~M.~Quang, C.~Liu, and W.~Xiang are with the School of Computing, Engineering, and Mathematical Sciences, La Trobe University, Melbourne, VIC 3086, Australia (e-mail: quang.nguyen@latrobe.edu.au; c.liu6@latrobe.edu.au; w.xiang@latrobe.edu.au).}%
\thanks{S.~Li is with the Chair of Communications and Information Theory, Technical University of Berlin, 10623 Berlin, Germany (e-mail: shuangyang.li@tu-berlin.de).}
\thanks{H.-N.~Vu is with the Faculty of Artificial Intelligence, Posts and Telecommunications Institute of Technology, Hanoi 12110, Vietnam (e-mail: namvh@ptit.edu.vn).}
\thanks{D. W. K. Ng is with the School of Electrical Engineering and Telecommunications, University of New South Wales, Sydney, NSW 2052, Australia (e-mail: w.k.ng@unsw.edu.au).}
}

\markboth{Journal of \LaTeX\ Class Files,~Vol.~14, No.~8, August~2021}%
{Shell \MakeLowercase{\textit{et al.}}: A Sample Article Using IEEEtran.cls for IEEE Journals}


\IEEEaftertitletext{\vspace{-2\baselineskip}}
\maketitle
\begin{abstract}
Recently, environment reconstruction (ER) in integrated sensing and communication (ISAC) systems has emerged as a promising approach for achieving high-resolution environmental perception. However, the initial results obtained from ISAC systems are coarse and often unsatisfactory due to the high sparsity of the point clouds and significant noise variance. To address this problem, we propose a noise–sparsity-aware diffusion model (NSADM) post-processing framework. Leveraging the powerful data recovery capabilities of diffusion models, the proposed scheme exploits spatial features and the additive nature of noise to enhance point cloud density and denoise the initial input. Simulation results demonstrate that the proposed method significantly outperforms existing model-based and deep learning-based approaches in terms of Chamfer distance and root mean square error. 
\end{abstract}


\section{Introduction}

Environment reconstruction (ER) is a key enabler for critical applications such as autonomous navigation and smart-city monitoring~\cite{nguyen2024deep, lin_environment_2024, lu_deep-learning-based_2024, bazzi2025isac}. To support these applications, Integrated Sensing and Communication (ISAC)~\cite{liu2019deep, liu2020location, liu2022scalable} has emerged as a promising technology for future 6G networks, unifying sensing and communication capabilities within a single system design. By leveraging ISAC, ER can be realized by processing the received echoes to estimate the locations of reflection points, each corresponding to a detected object surface in a specific direction. Aggregating these reflections yields a point cloud representing the surroundings without requiring dedicated sensing hardware.
Despite its potential, ISAC-enabled ER faces significant challenges, such as limited bandwidth and low received signal-to-noise ratio (SNR)~\cite{liu2020deeptransfer, liu2019deep, xie2020deep}. Consequently, the ISAC-enabled point clouds often suffer from missed detections and high noise variance, resulting in a significant decline in ER performance. Thus, the post-processing of ISAC-enabled point clouds plays a critical role in ER, garnering significant attention from both academia and industry~\cite{liu_survey_2022}.

Recent research has explored post-processing techniques to enhance ISAC-enabled ER in terms of accuracy and density~\cite{lxm2020deepresidual, xie2020unsupervised, yuan2020learning}. \textcolor{revised}{For instance, Bazzi et al.~\cite{bazzi2025isac} extract multipath components from channel state information and develop a two-stage reflection point optimization algorithm for 3D ER.} In another direction, Lin et al.~\cite{lin_environment_2024} introduce a framework that uses uplink signals from multiple users, along with a user selection algorithm, to improve the precision of the reflection point estimation. Meanwhile, Lu et al.~\cite{lu_deep-learning-based_2024} present a deep learning-based pipeline that employs two neural networks for 4D environmental reconstruction, significantly improving the density and structural coherence of the output point clouds. Despite their acceptable performance, these aforementioned methods tend to focus exclusively on either mitigating point cloud sparsity or suppressing additive noise, while in practice, both issues often coexist.

Recently, diffusion models have demonstrated strong data recovery capabilities across various domains~\cite{yang_diffusion_2023}. Notably, diffusion models possess inherent denoising and generative capabilities, making them highly effective for point cloud refinement. Motivated by these strengths, this letter proposes a diffusion model-based approach to enhance ER in ISAC systems. First, we formulate a generalized point cloud estimation problem for ISAC-enabled ER that jointly accounts for range estimation noise and missed detections by incorporating CRB-based noise variance and detection probability modeling to capture the statistical characteristics of ISAC point clouds. Second, we propose a noise-sparsity-aware diffusion model (NSADM) framework, where the forward process integrates CRB-guided noise injection and sparsity simulation to emulate realistic ISAC degradation. In the reverse process, the denoising network is conditioned on the CRB-based noise variance and detection probability matrices, leading to improved denoising and densification performance. Finally, extensive simulations demonstrate that the proposed method significantly improves both qualitative reconstruction and quantitative metrics, achieving lower Chamfer Distance (CD) and root mean squared error (RMSE) than existing model-based and learning-based baselines.

\section{System Model and Problem Formulation}
\label{system-model}

\begin{figure}
    \centering
    \includegraphics[width=1\linewidth]{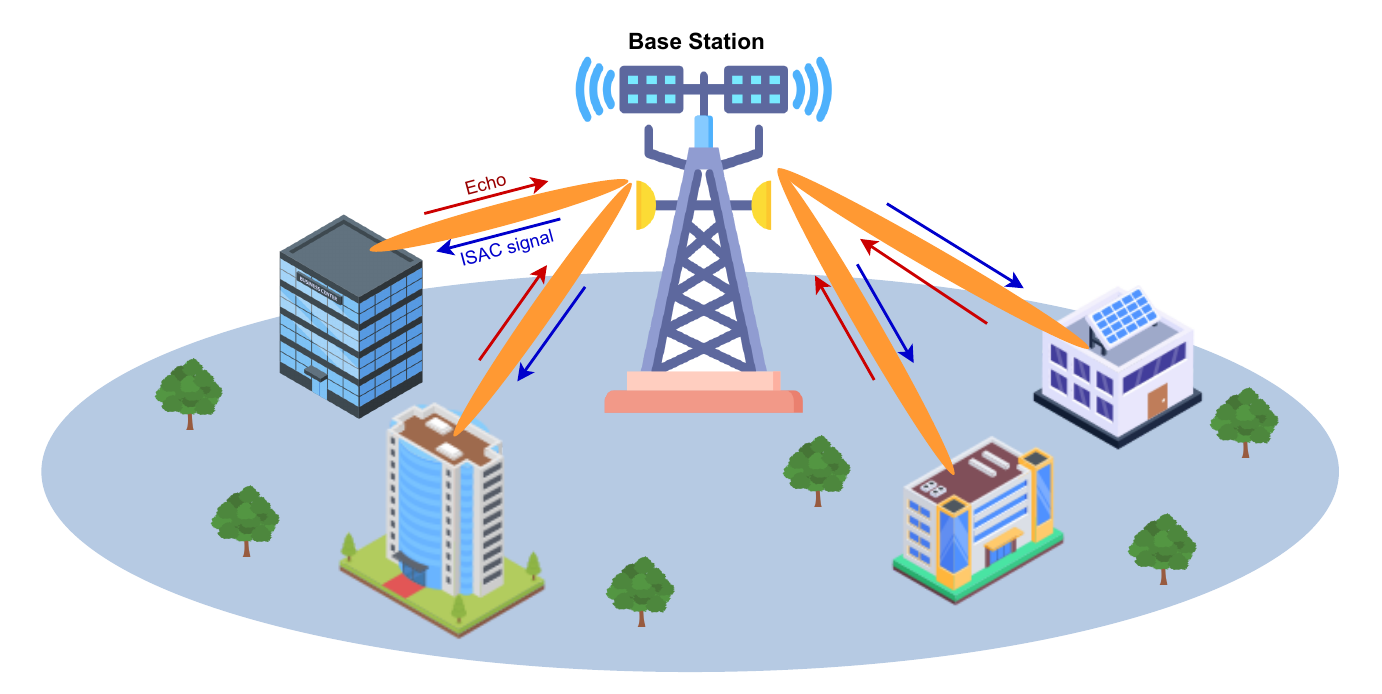}
    \caption{An illustration of the ISAC ER scenario.}\vspace{-15pt}
    \label{fig:target-scenario}
\end{figure}

As illustrated in Fig.~\ref{fig:target-scenario}, we consider a downlink ISAC scenario in which a base station (BS) transmits ISAC signals and collects reflected echoes from multiple static objects to conduct the task of ER. The BS is equipped with a square uniform planar array (UPA) consisting of \( N_t \times N_t \) transmit and \( N_r \times N_r \) receive antennas.

\subsection{Signal Model}
\label{sec:signal-model}

Let $s_n(t)$ denote the transmitted symbol at time instant $t$ in the $n$-th time slot, with total transmit power $P_t$. The BS transmits
\begin{equation}
    \mathbf{x}_n(t) = \mathbf{w}_n s_n(t),
\end{equation}
where $\mathbf{w}_n$ is the transmit beamforming vector. The beamforming vector is selected from a predefined codebook $\mathcal{B} = \left\{ \mathbf{a}(\phi_i,\theta_j) \,\middle|\, \phi_i \in \Phi,\; \theta_j \in \Theta \right\}$~\cite{lin_environment_2024}, where $\Phi = \{\phi_1,\dots,\phi_W\}$ and $\Theta = \{\theta_1,\dots,\theta_H\}$ denote the discrete sets of azimuth and elevation angles, with $W$ and $H$ representing the total number of azimuth and elevation grid points, respectively. At each time slot, the BS sets
$\mathbf{w}_n = \mathbf{a}(\phi_i,\theta_j)$,
where $\mathbf{a}(\phi_i,\theta_j) \in \mathbb{C}^{N_t^2 \times 1}$ is the steering vector toward direction $(\phi_i,\theta_j)$. For a square UPA with $N_t \times N_t$ elements and half wavelength spacing, the steering vector is written as
\begin{equation}
    \mathbf{a}(\phi_i,\theta_j)
    = \frac{1}{\sqrt{N_t^2}}
      \big[\, e^{j\pi\left(m\sin\theta_j\cos\phi_i + n\sin\theta_j\sin\phi_i\right)} \,\big]_{p,q=0}^{N_t-1},
\end{equation}
where each entry corresponds to the antenna element indexed by $(p,q)$. Under the line-of-sight (LoS) channel model\footnote{\textcolor{revised}{The outdoor mmWave propagation ensures that the LoS path dominates in our system, justifying the deterministic channel approximation. Future work will extend the framework to more practical settings with explicit Rician fading characterization for environments with varying effective K-factors.}}, the received echo at the BS corresponding to beam direction $(\phi_i,\theta_j)$ can be written as~\cite{liu2022learning}
\begin{equation}
\begin{aligned}
\mathbf{y}_{n,i,j}(t)
    = \, &\sqrt{N_t^2 N_r^2}\, \beta_{i,j} \mathbf{a}(\phi_i,\theta_j)\mathbf{a}^H(\phi_i,\theta_j) \mathbf{x}_n(t - \tau_{i,j}) \\
    &+ \mathbf{z}(t),
    \label{eq:received-signal}
\end{aligned}
\end{equation}
where $\tau_{i,j}$ is the round-trip delay, and $\beta_{i,j}=\frac{\varrho_{i,j}}{2d_{i,j}}$ is the reflection coefficient determined by the distance $d_{i,j}$ and radar cross section (RCS) $\varrho_{i,j}$. \textcolor{revised}{Here, $\varrho_{i,j}$ depends on the static object geometry, surface material, and incidence angle, whereas motion-induced variations are not considered under the quasi-static setup~\cite{lin_environment_2024}.
The term $\mathbf{z}(t) \sim \mathcal{CN}(\mathbf{0}, \sigma_z^2\mathbf{I}_{N_r^2})$ models aggregate sensing noise, including thermal noise and residual clutter returns from non-target scatterers\footnote{\textcolor{revised}{While our current framework focuses on dominant-path reconstruction, clutter returns could be exploited to enhance reconstruction completeness in future extensions.}}. In monostatic operation, self-interference is suppressed using a cancellation technique~\cite{ahmed2015all}, then the residual leakage is absorbed into $\mathbf{z}(t)$.}  Finally, only one beam is used per sensing slot, avoiding inter-beam interference.

\subsection{Environment Reconstruction Model}
\label{sec:reflection-estimate}

For each transmit beam direction $(\phi_i, \theta_j)$, the corresponding received signal $\mathbf{y}_{n,i,j}(t)$ is processed to estimate the round-trip time delay $\tau_{i,j}$ using conventional matched filtering approach~\cite{liu2022learning}:
\begin{equation}
    \tilde{\tau}_{i,j} = 
    \arg\max_{\tau} \left| \int_{0}^{\Delta T_e} \mathbf{y}_{n,i,j}(t)\, s_n^*(t - \tau)\, dt \right|^2,
\end{equation}
where $\Delta T_e$ denotes the observation window. The estimated distance is then 
$\tilde{d}_{i,j} = \frac{c\,\tilde{\tau}_{i,j}}{2} =d_{i, j} + \epsilon_{i, j}$,
where $c$ is the speed of light and $\epsilon_{i,j} \sim \mathcal{N}(0, (\sigma_{i,j}^d)^2)$ is estimation error of $d_{i,j}$ with noise variance $(\sigma_{i,j}^d)^2$. Repeating this procedure across all beam directions $(\phi_i, \theta_j)$ in $\mathcal{B}$ yields an estimated distance matrix $\tilde{\mathbf{\Omega}} = \{\tilde{d}_{i,j}\} \in \mathbb{R}^{W \times H}.$ This distance matrix (DM) can be mapped into 3D Cartesian coordinates to form a coarse point cloud expressed as
\begin{equation}
    \tilde{\mathbf{P}} = \{ \mathbf{p}_{i,j} \} \in \mathbb{R}^{3 \times W \times H}, \quad 
    \mathbf{p}_{i,j} = \tilde{d}_{i,j} 
    \begin{bmatrix}
        \cos\theta_j \cos\phi_i \\
        \cos\theta_j \sin\phi_i \\
        \sin\theta_j
    \end{bmatrix},
\end{equation}
which provides an initial reconstruction of the surrounding environment. However, the estimated point cloud from the sensing process suffers from two key limitations: high noise variance of distance estimation and {insufficient spatial sampling}. In particular, both are related to the low SNR, long distance, and limited bandwidth in ISAC systems. Based on the received signal model in Eq.~\eqref{eq:received-signal}, \textcolor{revised}{and assuming ideal beam alignment, under which each codebook beam achieves its maximum directional gain~\cite{liu2022learning}}, the SNR corresponding to $(\phi_i, \theta_j)$ is given by
\begin{equation}
\begin{aligned}
\gamma_{i,j} 
&= \frac{\mathbb{E}\!\left\{ \big\|\sqrt{N_t^{2}N_r^{2}}\;\beta_{i,j}\,
\mathbf{a}(\phi_i,\theta_j)\mathbf{a}^H(\phi_i,\theta_j)\,
\mathbf{x}_n(t-\tau_{i,j})\big\|^{2} \right\}}
{\mathbb{E}\!\left\{ \|\mathbf{z}(t)\|^{2} \right\}}
\\[0.5em]
&= \frac{N_t^{2}N_r^{2}\,|\beta_{i,j}|^{2}\,P_t}{\sigma_z^{2}}  =  \frac{N_t^{2}N_r^{2}\,\varrho_{i,j}^{2}\,P_t}{4\,d_{i,j}^{2}\,\sigma_z^{2}}.
\end{aligned}
\end{equation}
To characterize the uncertainty in the estimated distance \( \tilde{d}_{i,j} \), {we approximate the distance estimation error variance using the Cramér–Rao bound (CRB)~\cite{liu_survey_2022}, which sets a theoretical lower bound on the variance of unbiased estimators.} Then, the noise variance of the distance estimation error is given by~\cite{liu_survey_2022}
\begin{equation}
(\sigma_{i,j}^d)^2= \alpha \ \mathrm{CRB}_{i,j} = \frac{\alpha \ c^2}{8 \pi^2 \ \gamma_{i,j} \ \mathrm{B}^2},
\label{eq:crb}
\end{equation}
where \( \mathrm{B} \) is the effective bandwidth, and \(\alpha \geq 1\) is a scaling factor that accounts for estimator efficiency. In practice, the variance inflation factor is typically bounded as $1 \leq \alpha \leq 2,$ and seldom exceeds this range in radar and communication estimation problems~\cite{stoica2022CRBQuantized}. In addition to degraded accuracy, the point cloud suffers from missed detections, leading to sparsely populated reconstructions. The detection probability of reflection point corresponding to $(\phi_i, \theta_j)$ is derived by Marcum $Q$-function as follow~\cite{shnidman2002radar}
\begin{equation}
\begin{split}
    P_{i,j} &= Q_1\left(\sqrt{2\gamma_{i,j}},\ \sqrt{2\xi}\right) \\ 
    &= \int_{\sqrt{2\xi}}^{\infty} x \exp\left(-\frac{x^2 + 2\gamma_{i,j}}{2}\right) 
     I_0\left(x \sqrt{2\gamma_{i,j}}\right)\, dx,
     \label{eq:detection-probability}
\end{split}
\end{equation}
where \( I_0(\cdot) \) is the modified Bessel function of the first kind and zero order. Here, $\xi$ denotes the normalized detection threshold, which is chosen to satisfy a predefined false-alarm probability $P_\text{FA}$ under the Neyman–Pearson criterion, i.e., $\xi=-\ln P_\text{FA}$ for single noncoherent detection~\cite{shnidman2002radar}. 

{At the single-beam level, the estimate $\tilde{d}_{i,j}$ is affected by additive noise and potential missed detection, with variance approximated by the CRB in Eq.~\eqref{eq:crb} and detection probability $P_{i,j}$ given by the Marcum-$Q$ function in Eq.~\eqref{eq:detection-probability}. Extending this to the system level, when the BS performs beam sweeping over the codebook $\mathcal{B}$, these statistical characteristics are preserved across all directions $(\phi_i,\theta_j)$. Thus, the estimated DM $\tilde{\mathbf{\Omega}}$ inherits two degradation mechanisms: (i) Gaussian errors with variance matrix $\mathbf{\Sigma}_d = \mathrm{diag} \ \!\big( (\sigma^d_{1,1})^2,\, (\sigma^d_{1,2})^2,\, \ldots,\, (\sigma^d_{W,H})^2 \big)\;\in\;\mathbb{R}^{WH\times WH}$ and (ii) missed detections governed by the probability matrix $\mathbf{\Psi} = \{P_{i,j} \} \in \mathbb{R}^{W \times H}$. This yields the compact representation
\begin{equation}
\tilde{\mathbf{\Omega}} = ( \mathbf{\Omega} + \mathbf{\epsilon} ) \odot \mathbf{M},
\label{ER-model}
\end{equation}
where $\mathbf{\Omega}$ denotes the ground-truth DM, $\mathbf{\epsilon}\sim \mathcal{N}(0,\mathbf{\Sigma}_d)$ models CRB-based errors, and $\mathbf{M}=[M_{i,j}] \in \{0,1\}^{W \times H}$ is a random mask with $\Pr(M_{i,j}=1)=\mathbf{\Psi}_{i,j}$.}

\subsection{Problem Formulation}
In this letter, we investigate ER in ISAC, whose performance depends on the fidelity of the estimated DM $\tilde{\mathbf{\Omega}}$. Therefore, our objective is to design a mapping function $\mathcal{F}(\cdot)$ that refines the coarse and noisy estimation $\tilde{\mathbf{\Omega}}$ into a more accurate and dense DM $\widehat{\mathbf{\Omega}}$, which can then be used to reconstruct the 3D environment. Additionally, we incorporate the CRB-based noise variance matrix $\mathbf{\Sigma}_d$ and the detection probability matrix $\mathbf{\Psi}$ as auxiliary inputs. The resulting optimization problem is formulated as
\begin{equation}
\begin{aligned}
    &\min_{\mathcal{F}} \;\; \mathcal{L}\!\left(\widehat{\mathbf{\Omega}}, \mathbf{\Omega}\right), 
    \quad \widehat{\mathbf{\Omega}} = \mathcal{F}\!\left(\tilde{\mathbf{\Omega}}, [\mathbf{\Psi}, \mathbf{\Sigma}_d]\right) \\
    &\text{s.t.}\;\; 
    \tfrac{1}{WH}\,\|\mathbf{\Psi}\|_{1} \;\;\geq\; \rho_0, \quad \tfrac{1}{WH}\,\mathrm{tr}(\mathbf{\Sigma}_d) \;\leq\; \sigma_0^2,
\end{aligned}
\label{eq:problem-formulation}
\end{equation}
where $\mathcal{L}(\cdot)$ measures the geometric discrepancy between the predicted and ground-truth distance matrices. To ensure realistic reconstruction conditions, we impose a minimum average detection probability $\rho_0$ to guarantee sufficient geometric coverage and an upper bound on the average noise variance $\sigma_0^2$ to limit estimator uncertainty. Eq.~\eqref{eq:problem-formulation} constitutes a high-dimensional and ill-posed inverse problem~\cite{xie2019activity, liu2019maximum}. To address this challenge, we propose a {NSADM} framework that leverages the CRB-based noise variance matrix and the detection probability matrix to guide the reconstruction process.

\begin{figure*}
    \centering
    \vspace{-10pt}
    \includegraphics[width=0.9\linewidth]{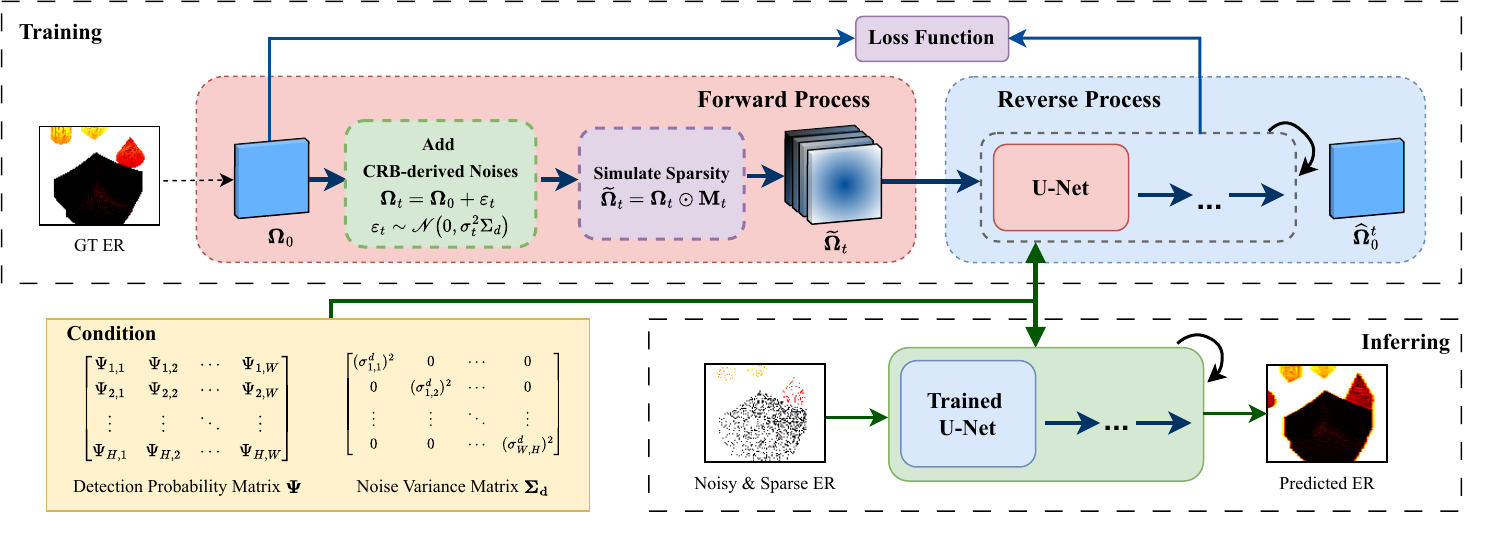}\vspace{-5pt}
    \caption{\textcolor{revised}{Overview of the proposed NSADM framework.}}\vspace{-17pt}
    \label{fig:architecture-overview}
\end{figure*}

\section{NSADM: Noise-Sparsity-Aware Diffusion Model-based Point Cloud Enhancement}


Fig.~\ref{fig:architecture-overview} illustrates the overall architecture of the NSADM framework. During training, the forward process sequentially injects CRB-based anisotropic noise and simulates beam-dependent sparsity. The reverse process then exploits the noise variance matrix $\mathbf{\Sigma}_d$ and detection probability matrix $\mathbf{\Psi}$ as conditioning information to denoise and densify the corrupted DM. The details of the proposed framework are discussed in the following subsections.

\subsection{Noise-Sparsity-Aware Forward Process}
\label{sec:forward-process}
\textcolor{revised}{Traditional denoising diffusion probabilistic models (DDPM)~\cite{yang_diffusion_2023} inject isotropic Gaussian noise with an identity covariance, which does not align with the anisotropic noise characteristics of ISAC measurements due to varying propagation distances and target RCS. Therefore, we adopt a score-based generative model (SGM)~\cite{yang_diffusion_2023} which enables customizable noise covariance and flexible noise scheduling, allowing the diffusion process to more accurately reflect the physical noise statistics of ISAC systems.} Particularly, the forward process perturbs the ground-truth DM $\mathbf{\Omega}_0$ with anisotropic Gaussian noise using a normalized CRB-based covariance matrix $\widehat{\mathbf{\Sigma}}_d$ under increasing noise levels $0 < \sigma_1 < \sigma_2 < \dots < \sigma_T$. At timestep $t$, the corrupted sample is drawn as
\begin{equation}
    q(\mathbf{\Omega}_t \mid \mathbf{\Omega}_{0}) 
    = \mathcal{N}\!\left(\mathbf{\Omega}_t ; \mathbf{\Omega}_{0}, \, \sigma_t \widehat{\mathbf{\Sigma}}_d\right), \quad \widehat{\mathbf{\Sigma}}_d = \frac{W H}{\operatorname{tr}(\mathbf{\Sigma}_d)}\,\mathbf{\Sigma}_d
\end{equation}
where $\sigma_t$ is the noise level at timestep $t$.

As discussed in Sec.~\ref{sec:reflection-estimate}, in addition to noise, ISAC reconstructions suffer from sparse detections. To model this, we construct a detection probability matrix $\mathbf{\Psi} \in [0,1]^{W \times H}$, where $\mathbf{\Psi}_{i,j}$ denotes the probability of detecting a reflection in beam direction $(\phi_i,\theta_j)$ as defined in Sec.~\ref{sec:reflection-estimate}. Based on $\mathbf{\Psi}$, we generate a binary sampling mask $\mathbf{M}_t \in \{0,1\}^{W \times H}$ following a sampling schedule $\rho_t \in [0,1]$ that controls the proportion of retained points at step $t$:
\begin{equation}
\mathbf{M}_{t,i,j} = 
\begin{cases}
1, & \text{if } \rho_t \mathbf{\Psi}_{i,j} > \mu, \quad \mu \sim \mathcal{U}(0, 1), \\
0, & \text{otherwise}.
\end{cases}
\end{equation}
Applying this mask to the noisy DM yields
$\tilde{\mathbf{\Omega}}_t = \mathbf{M}_t \odot \mathbf{\Omega}_t$.
This step enforces sparsity consistent with ISAC detection statistics and produces the degraded input $\tilde{\mathbf{\Omega}}_t$ used by the reverse denoising process.

\subsection{Noise-Sparsity-Conditioned Reverse Process}
In SGMs, the reverse process iteratively denoises the corrupted sample by approximating the posterior transition $p(\mathbf{\Omega}_{t-1}\,|\, {\mathbf{\Omega}}_t)$. We augment this process by conditioning on $\mathbf{C}=\{\mathbf{\Psi},\widehat{\mathbf{\Sigma}}_d\}$. This conditioning reduces posterior uncertainty, as quantified by conditional mutual information:
\begin{equation}
\mathrm{I}(\mathbf{\Omega};\mathbf{C}\mid \tilde{\mathbf{\Omega}}) = \mathrm{H}(\mathbf{\Omega} \mid \tilde{\mathbf{\Omega}}) - \mathrm{H}(\mathbf{\Omega}\mid \tilde{\mathbf{\Omega}},\mathbf{C}) \;\ge\; 0,
\end{equation}
where $\mathrm{H}(\cdot)$ denotes Shannon entropy.
Following the Karras formulation~\cite{karras_elucidating_2022}, at each timestep $t$, we employ a denoising network $R_{\mathrm{w}}(\cdot)$, parameterized by weights $\mathrm{w}$, to directly estimate the clean distance matrix:
\begin{equation}
    \widehat{\mathbf{\Omega}}_{0}^{t} = R_{\mathrm{w}}(\mathbf{\Omega}_t, t, \mathbf{C}).
    \label{eq:clean_prediction}
\end{equation}
The conditional reverse transition $p_{\mathrm{w}}$ can be defined as
\begin{equation}
    p_{\mathrm{w}}(\mathbf{\Omega}_{t-1}\mid\mathbf{\Omega}_t, \mathbf{C}) 
    = \mathcal{N}\!\left(\mathbf{\Omega}_{t-1}; \widehat{\mathbf{\Omega}}_{0}^{t}, \, \sigma_t^2 \widehat{\mathbf{\Sigma}}_d\right),
    \label{eq:reverse_transition}
\end{equation}
where $\widehat{\mathbf{\Omega}}_{0}^{t}$ serves as the conditional mean. This formulation enables iterative refinement from the degraded DM $\mathbf{\Omega}_T$ to the predicted clean DM $\widehat{\mathbf{\Omega}}_0$.

\subsection{Objective Function}

Training the NSADM framework is guided by a hybrid loss that balances reconstruction fidelity with structural consistency. The total loss is defined as
\begin{equation}
\mathcal{L}_\text{total} =
\underbrace{\left\| \mathbf{\Omega}_0 - \widehat{\mathbf{\Omega}}_{0}^{t} \right\|_2^2}_{\text{MSE Loss}} +
\lambda \underbrace{\left\| f(\mathbf{\Omega}_0) - f\!\left(\widehat{\mathbf{\Omega}}_{0}^{t}\right) \right\|_2^2}_{\text{Perceptual Loss}},
\label{eq:loss-function}
\end{equation}
\textcolor{revised}{where $f(\cdot)$ is a fixed feature extractor which is pretrained to capture structural patterns (e.g., edges and contours) that are also present in ISAC distance matrices,} and $\lambda$ controls the trade-off between fidelity and perceptual quality.

The first term is a mean squared error (MSE) loss, which penalizes deviations in per-beam distance estimation in regions where reliable measurements exist. However, MSE operates solely in pixel space and tends to oversmooth reconstructions, particularly in sparse or noisy areas. \textcolor{revised}{To complement this, we incorporate a perceptual loss based on the learned perceptual image patch similarity (LPIPS) metric~\cite{zhang2018unreasonable}, which leverages $f(\cdot)$ to preserve structural consistency in distance matrices that cannot be enforced by MSE alone.}

\subsection{Training and Inference}

During training, the model learns to predict the ground-truth DM at each diffusion step guiding by the hybrid loss in Eq.~\eqref{eq:loss-function}. At inference, the model takes the degraded DM $\tilde{\mathbf{\Omega}}$ and applies the conditional reverse process in Eq.~\eqref{eq:reverse_transition}. Each step refines the estimate by denoising and densifying under the guidance of $\mathbf{\Psi}$ and $\widehat{\mathbf{\Sigma}}_d$. After $T$ iterations, the final DM $\widehat{\mathbf{\Omega}}$ is mapped to 3D Cartesian coordinates to produce the predicted point cloud $\widehat{\mathbf{P}}$.

\textcolor{revised}{The computational complexity of the proposed model follows the standard diffusion models in~\cite{karras_elucidating_2022}. Particularly, each denoising step is dominated by convolutional operations over a feature map of spatial dimensions $N_h \times N_w$ with $N_c$ channels, leading to a per-step cost of $\mathcal{O}(N_c N_h N_w)$. Therefore, the overall inference complexity scales linearly with the number of sampling steps $T$, resulting in a total cost of $\mathcal{O}(T N_c N_h N_w)$.}

\begin{figure*}[htbp]
\centering
\renewcommand{\arraystretch}{1.0}
\setlength{\tabcolsep}{3pt}

\begin{tabular}{c c c c c c}
\small \textbf{Ground Truth} & \small \textbf{Initial Estimation} & \small \textbf{Proposed NSADM} & \small \textbf{DPIR} & \small \textbf{DnCNN} & \small \textbf{MT} \\
\raisebox{-0.5\height}{\includegraphics[width=0.15\textwidth]{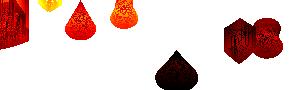}} &
\raisebox{-0.5\height}{\includegraphics[width=0.15\textwidth]{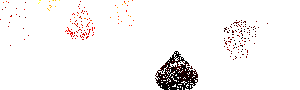}} &
\raisebox{-0.5\height}{\includegraphics[width=0.15\textwidth]{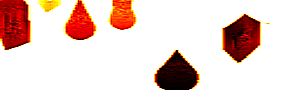}} &
\raisebox{-0.5\height}{\includegraphics[width=0.15\textwidth]{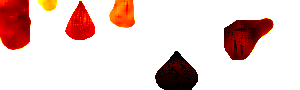}} &
\raisebox{-0.5\height}{\includegraphics[width=0.15\textwidth]{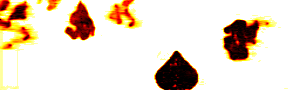}} &
\raisebox{-0.5\height}{\includegraphics[width=0.15\textwidth]{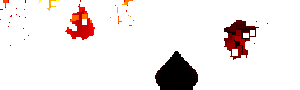}} \\
\raisebox{-0.5\height}{\includegraphics[width=0.15\textwidth]{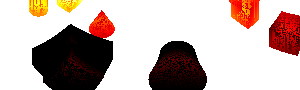}} &
\raisebox{-0.5\height}{\includegraphics[width=0.15\textwidth]{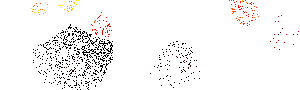}} &
\raisebox{-0.5\height}{\includegraphics[width=0.15\textwidth]{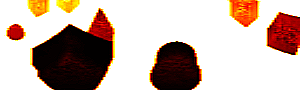}} &
\raisebox{-0.5\height}{\includegraphics[width=0.15\textwidth]{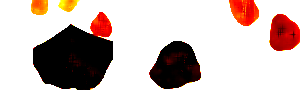}} &
\raisebox{-0.5\height}{\includegraphics[width=0.15\textwidth]{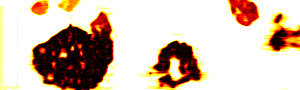}} &
\raisebox{-0.5\height}{\includegraphics[width=0.15\textwidth]{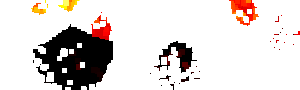}} \\
\end{tabular}

\caption{Qualitative comparison of point cloud reconstruction across two representative scenes. From left to right: Ground Truth, initial ISAC estimation, Proposed NSADM, DPIR~\cite{zhang2021plug}, DnCNN~\cite{zhang2017beyond}, and MT~\cite{sreedhar2012enhancement}. The darker colors indicate closer distances to the BS, while lighter colors represent farther points.}\vspace{-10pt}
\label{fig:vis}
\end{figure*}

\section{Experimental Results}
\subsection{Dataset Generation and Benchmark Selection}
\textcolor{revised}{To evaluate the performance of the proposed framework, we generate a synthetic ISAC dataset of 1{,}000 outdoor-like scenes using a simulation pipeline implemented in \texttt{Python3.11} with standard scientific libraries. Each scene is constructed using a 3D mesh engine, where multiple objects with varying shapes and RCS values are randomly placed within a 30 m sensing radius. A monostatic BS, positioned at a height of 20–25 m and equipped with a $4\times4$ UPA, scans over a predefined azimuth–elevation codebook. The system operates with a 1 GHz bandwidth and a noise power of $-20$ dBm. Echo generation and range estimation follow the signal and processing models in Sec.\ref{sec:signal-model} and Sec.\ref{sec:reflection-estimate}, acquiring noisy distance matrices that are then mapped to Cartesian coordinates to form point clouds. The dataset is then partitioned into training, validation, and testing subsets. All experiments are performed on an NVIDIA GeForce RTX3090 GPU with an AMD EPYC~7313 16-core processor. Benchmark comparisons include DPIR~\cite{zhang2021plug},  DnCNN~\cite{zhang2017beyond}, and morphological transformation (MT)~\cite{sreedhar2012enhancement}, representing diffusion-prior-based methods, discriminative learning-based, and classical model-based respectively.}

\subsection{Performance Evaluation}

\textit{1) Qualitative Comparison:} Fig.~\ref{fig:vis} presents a qualitative comparison, showing that the proposed NSADM framework outperforms all baselines by jointly achieving effective densification and strong denoising, producing clear object boundaries and structurally consistent scenes. In contrast, DnCNN only partially suppresses noise and fails to recover missing structures, producing fragmented reconstructions due to its non-iterative processing. \textcolor{revised}{Meanwhile, DPIR offers stronger noise suppression by leveraging diffusion priors and iterative refinement, yielding visually sharper reconstructions. However, the lack of perceptual loss leads to weak geometric fidelity, with noticeable shape deformation such as rounded corners and misaligned edges.} MT, while able to interpolate gaps when neighbors exist, lacks denoising capability and fails to reconstruct in highly sparse areas. Under very low SNR, the high missed detection probability causes all methods, including NSADM, to produce incomplete reconstructions or spurious objects. Addressing this fundamental challenge is a key direction for future work.

\begin{figure}[t]
  \centering
  \begin{tabular}{c c}
    \includegraphics[width=0.48\linewidth] {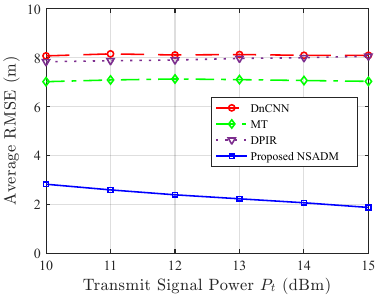} \hspace{-10pt} &
    \includegraphics[width=0.48\linewidth]{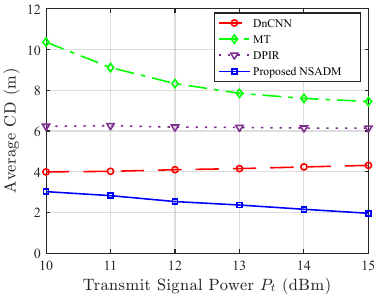} \\[-5pt]
    \small (a) &
    \small (b)
  \end{tabular}
  \vspace{-5pt}
  \caption{Performance comparison of NSADM, DPIR, DnCNN, and MT in terms of (a) average RMSE and (b) average CD across varying transmit powers.}
    \vspace{-5pt}
  \label{fig:method_comparison}
\end{figure}

\begin{figure}[t]
    \centering
    \includegraphics[width=0.65\linewidth]{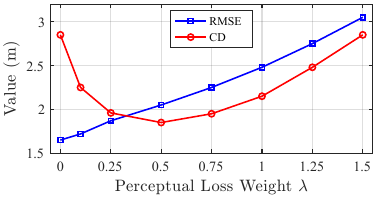}
    \caption{\textcolor{revised}{Trade-off between fidelity and perceptual quality. Varying perceptual loss weight $\lambda$ in the hybrid loss function affects average RMSE and CD.}}
    \vspace{-10pt}
    \label{fig:tradeoff}
\end{figure}

\textit{2) Quantitative Comparison:} \textcolor{revised}{We adopt RMSE and CD as quantitative metrics for evaluating reconstruction performance. While RMSE corresponds to the MSE loss, perceptual loss and CD share an emphasis on structural coherence. The structural patterns preserved by perceptual loss may improve CD performance through the deterministic spherical-to-Cartesian mapping. Particularly, perceptual loss preserves geometric patterns (e.g., sharp boundaries, smooth surfaces)~\cite{zhang2018unreasonable} in the distance matrix during training, while CD evaluates geometric alignment in the reconstructed point cloud during evaluation~\cite{bazzi2025isac}. The CD is given by
\begin{equation}
    \mathrm{CD}(\widehat{\mathbf{P}}, \mathbf{P}) 
    = \tfrac{1}{|\widehat{\mathbf{P}}|}\!\sum_{x\in \widehat{\mathbf{P}}} \min_{y\in \mathbf{P}} \|x-y\|_2^2
    + \tfrac{1}{|\mathbf{P}|}\!\sum_{y\in \mathbf{P}} \min_{x\in \widehat{\mathbf{P}}} \|y-x\|_2^2.
\label{eq:chamfer-distance}
\end{equation}
Specifically, CD aggregates bidirectional nearest-neighbor distances, making it sensitive to both missing regions and spurious detections, which is appropriate for environment reconstruction where geometric completeness is critical.}
Fig.~\ref{fig:method_comparison} reports average RMSE and CD across all test scenes under varying transmit powers. NSADM achieves the lowest errors across all conditions, reducing RMSE by over 60\% and CD by over 50\% compared to the strongest baseline, which is MT and DnCNN, respectively. DnCNN exhibits nearly constant performance across different powers, reflecting its limited sensitivity to SNR and its denoising mechanism. Conversely, MT improves slightly as power increases but remains unreliable in sparse regions due to its dependence on neighboring reflections. \textcolor{revised}{Moreover, DPIR performs better than DnCNN in relatively dense areas owing to its diffusion priors, yet it still introduces geometric distortions due to the lack of ISAC-specific physical constraints.} In contrast, NSADM's errors decrease consistently with transmit power, demonstrating the effectiveness of its noise-sparsity-aware formulation.

\balance
\vspace{3pt}
\textcolor{revised}{\textit{3) Trade-off between fidelity and perceptual quality:} Fig.~5 examines the impact of perceptual loss weight $\lambda$ on reconstruction metrics. As $\lambda$ increases from 0 to 1.5, RMSE increases approximately linearly, indicating a trade-off between perceptual emphasis and point-wise fidelity. In contrast, CD exhibits distinct behavior across two regimes: from $\lambda = 0$ to $\lambda = 0.5$, CD decreases substantially with particularly sharp improvement from 0 to 0.25, demonstrating that incorporating perceptual loss significantly enhances geometric quality with modest RMSE cost; for $\lambda > 0.5$, CD increases progressively as excessive perceptual weighting introduces over-regularization that prioritizes structural coherence over geometric accuracy. These results validate $\lambda = 0.25$ as achieving optimal balance.}

\section{Conclusion}
This letter presented NSADM, a diffusion-based framework for refining sparse and noisy ISAC-enabled point clouds. By modeling CRB-based noise and detection sparsity, the method aligns the denoising process with ISAC-specific degradations, yielding denser and more structurally accurate reconstructions. Experiments across varying transmit powers demonstrate consistent gains over learning-based and interpolation baselines. \textcolor{revised}{Future work will extend the framework to dynamic scenarios and explore adaptive sensing strategies.}

\bibliographystyle{IEEEtran}
\bibliography{short_refs}

\end{document}